\documentclass[preprint,aps,amsmath,showpacs,showkeys,nofootinbib, 
superscriptaddress]{revtex4} 
\usepackage[dvips]{graphicx} 
 
\begin{document} 

\title{\textbf{Relativistic three-body calculations of 
a $Y=1,I=\frac{3}{2},J^P=2^+$ \\ $\pi\Lambda N-\pi\Sigma N$ dibaryon}} 
\author{H.~Garcilazo} 
\email{humberto@esfm.ipn.mx} 
\affiliation{Escuela Superior de F\' \i sica y Matem\'aticas \\ 
Instituto Polit\'ecnico Nacional, Edificio 9, 
07738 M\'exico D.F., Mexico} 
\author{A.~Gal} 
\email{avragal@savion.huji.ac.il} 
\affiliation{Racah Institute of Physics, The Hebrew University, 
Jerusalem 91904, Israel} 

\date{\today} 

\begin{abstract} 

The $\pi\Lambda N-\pi\Sigma N$ coupled-channel system with quantum numbers 
$(Y,I,J^P)=(1,\frac{3}{2},2^+)$ is studied in a relativistic three-body 
model, using two-body separable interactions in the dominant $p$-wave 
pion-baryon and $^3S_1$ $YN$ channels. Three-body equations are solved in 
the complex energy plane to search for quasibound-state and resonance poles, 
producing a robust narrow $\pi\Lambda N$ resonance about 10--20 MeV below 
the $\pi\Sigma N$ threshold. Viewed as a dibaryon, it is a $^{5}S_2$ 
quasibound state consisting of $\Sigma(1385)N$ and $\Delta(1232)Y$ 
components. Comparison is made between the present relativistic model 
calculation and a previous, outdated nonrelativistic calculation which 
resulted in a $\pi\Lambda N$ bound state. Effects of adding a $\bar K NN$ 
channel are studied and found insignificant. Possible production and decay 
reactions of this $(Y,I,J^P)=(1,\frac{3}{2},2^+)$ dibaryon are discussed. 

\end{abstract} 

%14.20.Pt: Exotic baryons   
%13.75.Gx: Pion-baryon interactions 
%13.75.Ev: Hyperon-nucleon interactions 
%11.80.Jy: Many-body scattering and Faddeev equation 


\maketitle 

\newpage

\section{Introduction} 
\label{sec:intro} 

In recent work \cite{GAGA08,GAGA10} we have studied the $\pi\Lambda N-
\pi\Sigma N$ coupled channel system, in which the dominant two-body 
configurations are the pion-nucleon $p$-wave $\Delta(1232)$ resonance with 
$s$-wave hyperon spectator, the pion-hyperon $p$-wave $\Sigma(1385)$ resonance 
with $s$-wave nucleon spectator, and the $YN$ ($Y\equiv\Lambda,\Sigma$) 
$^3S_1$ coupled channels with $p$-wave pion spectator. The contributions of 
these two-body configurations obviously maximize in the three-body channel 
with $(I,J^P)=(\frac{3}{2},2^+)$, where $I,J,P$ denote the total isospin, 
total angular momentum and parity, respectively. Substantial attraction in 
this three-body configuration was found in a nonrelativistic three-body 
calculation, resulting in a possible $\pi\Lambda N$ bound state. 
Having presented very recently a relativistic three-body Faddeev formalism 
appropriate for systems with $p$-wave two-body interactions \cite{GAGA11}, 
it is natural to apply it to the $\pi\Lambda N-\pi\Sigma N$ coupled channels 
system with $I=3/2$ and $J^P=2^+$. The main consequence of adopting 
a relativistic formalism, as shown below, is that the $\pi\Lambda N$ bound 
state dissolves, becoming a $\pi\Lambda N$ resonance below the $\pi\Sigma N$ 
threshold. We note that a relativistic three-body formalism equivalent to that 
of Ref.~\cite{GAGA11} was already applied in the context of searching for a 
$\bar KNN$ $(I=1/2,J^P=0^-)$ quasibound state for which the dominant two-body 
configurations are all $s$-waves \cite{Ikeda07}. We have also studied the 
effect of adding to the $(\frac{3}{2},2^+)$ $\pi YN$ channels a $\bar KNN$ 
channel, induced through a $\Sigma(1385)$-mediated two-body $p$-wave 
$\bar K N-\pi Y$ coupling, and found it to be relatively insignificant. 
This is to be expected, observing that none of the Pauli-allowed $s$-wave 
$NN$ configurations fits into a $(\frac{3}{2},2^+)$ $\bar KNN$ channel with 
a $p$-wave meson spectator. For a recent overview of dibaryon candidates and 
related studies, see Refs.~\cite{GAL11,Oka12,HM12}. 

The paper is organized as follows: input two-body interactions are described 
in Sect.~\ref{sec:2body} and three-body equations are derived in 
Sect.~\ref{sec:3body}. Results are described in Sect.~\ref{sec:res} and 
discussed in Sect.~\ref{sec:disc}. Several production reactions by which to 
search for the present $(Y,I,J^P)=(1,\frac{3}{2},2^+)$ dibaryon candidate 
are listed and briefly discussed in the Summary Sect.~\ref{sec:sum}.  

\newpage 

\section{Two-body interactions} 
\label{sec:2body} 

As discussed in Ref.~\cite{GAGA08}, the dominant two-body interactions are in 
the $p$-wave $\pi N$ $(I,J^P)=(\frac{3}{2},\frac{3}{2}^+)$ $\Delta(1232)$ and 
$\pi\Lambda-\pi\Sigma$ $(I,J^P)=(1,\frac{3}{2}^+)$ $\Sigma(1385)$ channels, 
and in the $s$-wave $\Lambda N-\Sigma N$ ($I=\frac{1}{2},{^{3}S_1}$) channel. 
We note that these two-body interactions, taken here in separable form, are 
independent of energy whereas the resulting two-body amplitudes are obviously 
energy dependent, and even resonate in the $p$-wave channels. 
Since the introduction of two-body {\it energy-dependent} interactions geared 
to simulate additional energy-dependent background amplitudes poses problems 
within a relativistic kinematics treatment (see Ref.~\cite{Ikeda10} for 
a recent discussion) we limit the two-body interaction input in the present 
three-body relativistic calculation to energy-independent separable forms 
described below. Our notational convention is to assign particle indices 
1,2,3 to hyperons, nucleon and pion, respectively.

\subsection{The $\pi N$ subsystem} 

The Lippmann-Schwinger equation for the pion-nucleon interaction is given 
by \cite{GAGA11}: 
\begin{eqnarray} 
t_1(p_1,p_1^\prime;\omega_0) &=& V_1(p_1,p_1^\prime)+\int_0^\infty 
{p_1^{\prime\prime}}^2 dp_1^{\prime\prime}  \nonumber \\ 
& \times & V_1(p_1,p_1^{\prime\prime})\frac{1}{\omega_0-
\sqrt{m_N^2+{p_1^{\prime\prime 2}}}-\sqrt{m_\pi^2+{p_1^{\prime\prime 2}}}+
{\rm i}\epsilon}t_1(p_1^{\prime\prime},p_1^\prime;\omega_0), 
\label{eq:A1} 
\end{eqnarray} 
so that using the separable potential
\begin{equation} 
V_1(p_1,p_1^\prime)=\gamma_1 g_1(p_1) g_1(p_1^\prime), 
\label{eq:A2} 
\end{equation} 
one gets 
\begin{equation}
t_1(p_1,p_1^\prime;\omega_0)= g_1(p_1) \tau_1(\omega_0) g_1(p_1^\prime), 
\label{eq:A3} 
\end{equation} 
where 
\begin{equation} 
[\tau_1(\omega_0)]^{-1} =\frac{1}{\gamma_1}-\int_0^\infty p_1^2 dp_1 
\frac{g_1^2(p_1)}{\omega_0-\sqrt{m_N^2+p_1^2}-\sqrt{m_\pi^2+p_1^2}+
{\rm i}\epsilon}. 
\label{eq:A4} 
\end{equation} 
A fit to the $P_{33}$ phase shift and scattering volume using the form factor 
\begin{equation} 
g_1(p_1)=p_1[\exp(-p_1^2/\beta_1^2)+Cp_1^2\exp(-p_1^2/\alpha_1^2)],
\label{eq:A5} 
\end{equation} 
and a set of parameters listed in Table~\ref{tab:piN}, row marked $P_{33}$, 
was shown and discussed in Ref.~\cite{GAGA11}. This form factor and parameters 
are used in the present calculations. Listed in the same row are also r.m.s. 
radii values of momentum-space and coordinate-space representations of the 
$P_{33}$ form factor. These were discussed too in Ref.~\cite{GAGA11}; here 
we recall that $\tilde{g_1}(r)$, the coordinate-space Fourier transform of 
$g_1(p)$, is not necessarily a nodeless function at finite values of $r$, 
so that an appropriate measure of its spatial extension is provided by the 
value of its (single) zero $r_0^{(\pi N)}$, given by the last entry. 
This does not appear to present a problem in the case of the $\pi N$ 
$P_{33}$ form factor, where the difference between the listed values of 
$\sqrt{<r^2>_{{\tilde{g_1}}}}$ and $r_0^{(\pi N)}$ is small, but it does 
present a problem in the case of the $\pi Y$ form factor where the squared 
radius $<r^2>_{{\tilde{g_1}}}$ assumes occasionally negative values. 
Returning to Table~\ref{tab:piN}, listed in the row marked $P_{13}$ are 
parameters fitted to the $P_{13}$ phase shifts which are considerably smaller 
than the $P_{33}$ resonating phase shifts. This $\pi N$ $P_{13}$ channel will 
act in the three-body calculation only together with a spectator $\Sigma$ 
hyperon, and its inclusion serves the purpose of estimating the role of 
$\pi B$ channels other than the resonating ones. For notational simplicity, 
and since the $\pi N$ $P_{13}$ channel is excluded from most of the 
calculations reported here, it is suppressed in the derivation of the 
three-body equations below. 

\begin{table}[ht]  
\caption{Fitted parameters of the $\pi N$ separable $p$-wave interaction 
(\ref{eq:A2}) with form factor $g_1(p)$ (\ref{eq:A5}). Values of the 
r.m.s. momentum $\sqrt{<p^2>_{g_1}}$ (fm$^{-1}$), r.m.s. radius 
$\sqrt{<r^2>_{{\tilde {g_1}}}}$ and zero $r_0^{(\pi N)}$ (both in fm) 
of the Fourier transform $\tilde {g_1}(r)$ are listed for the dominant 
$P_{33}$ channel.} 
\begin{ruledtabular} 
\begin{tabular}{cccccccc} 
channel & $\gamma_1~({\rm fm}^4)$ & $\alpha_1~({\rm fm}^{-1})$ & 
$\beta_1~({\rm fm}^{-1})$ & $C~({\rm fm}^2)$ & 
$\sqrt{<p^2>_{g_1}}$ & $\sqrt{<r^2>_{{\tilde {g_1}}}}$ & $r_0^{(\pi N)}$ \\ 
\hline 
$P_{33}$ & $-$0.075869 & 2.3668 & 1.04 & 0.23 & 4.07 & 1.47 & 1.36 \\ 
$P_{13}$ & 0.033       &   --   & 1.325 & 0.0 &      &      &      \\ 
\end{tabular} 
\end{ruledtabular} 
\label{tab:piN} 
\end{table} 

The $\pi N$ $P_{33}$ amplitude in the three-body system can have either 
$\Lambda$ or $\Sigma$ hyperon as spectator and is given by 
\begin{equation} 
t_1^Y(p_1,p_1^\prime;W_0,q_1)=g_1(p_1)\tau_1^Y(W_0,q_1) g_1(p_1^\prime), 
\label{eq:A6} 
\end{equation} 
where $W_0$ is the invariant mass of the three-body system, $q_1$ is the 
relative momentum between the hyperon and the c.m. of the $\pi N$ subsystem 
and 
\begin{equation} 
[\tau_1^Y(W_0,q_1)]^{-1} = \frac{1}{\gamma_1}-\int_0^\infty p_1^2 dp_1 
\frac{g_1^2(p_1)}{W_0-\sqrt{\left(\sqrt{m_N^2+p_1^2}+\sqrt{m_\pi^2+p_1^2}
\right)^2+q_1^2}-\sqrt{m_Y^2+q_1^2}+{\rm i}\epsilon}, 
\label{eq:A7} 
\end{equation} 
where $Y$ is either $\Lambda$ or $\Sigma$.

\subsection{The $\pi\Lambda - \pi\Sigma$ subsystem} 
\label{sec:piY} 

Since we have in this case two coupled channels the corresponding 
Lippmann-Schwinger equation is 
\begin{eqnarray} 
t_2^{YY'}(p_2,p_2^\prime;\omega_0) &=& V_2^{YY'}(p_2,p_2^\prime)+\sum_{Y''} 
\int_0^\infty {p_2^{\prime\prime}}^2 dp_2^{\prime\prime}  \nonumber \\ 
& \times & V_2^{YY''}(p_2,p_2^{\prime\prime})\frac{1}{\omega_0-\sqrt{m_{\pi}^2
+{p_2^{\prime\prime 2}}}-\sqrt{m_{Y''}^2+{p_2^{\prime\prime 2}}}+
{\rm i}\epsilon}t_2^{Y''Y'}(p_2^{\prime\prime},p_2^\prime;\omega_0). 
\label{eq:B1} 
\end{eqnarray} 
Here we used the separable potential 
\begin{equation} 
V_2^{YY'}(p_2,p_2^\prime)=\gamma_2 g_2^Y(p_2) g_2^{Y'}(p_2^\prime), 
\label{eq:B2} 
\end{equation} 
so that the solution of the Lippmann-Schwinger equation is 
\begin{equation} 
t_2^{YY'}(p_2,p_2^\prime;\omega_0)=
g_2^Y(p_2)\tau_2(\omega_0)g_2^{Y'}(p_2^\prime), 
\label{eq:B3} 
\end{equation} 
with 
\begin{equation} 
\tau_2^{-1}(\omega_0)=\frac{1}{\gamma_2}-\sum_Y \int_0^\infty p_2^2 dp_2 
\frac{[g_2^Y(p_2)]^2}{\omega_0-\sqrt{m_{\pi}^2+p_2^2}-\sqrt{m_Y^2+p_2^2}+
{\rm i}\epsilon}. 
\label{eq:B4} 
\end{equation} 
The two-body amplitude in the three-body system with a nucleon as spectator 
is given by expressions analogous to (\ref{eq:A6}) and (\ref{eq:A7}). 
Following Ref.~\cite{GAGA11} we used the form factors 
\begin{equation} 
g_2^{\Lambda}(p_2)=p_2(1+Ap_2^2)\exp(-p_2^2/\beta_2^2),\,\,\,\,\,\, 
g_2^{\Sigma}(p_2)=Bg_2^{\Lambda}(p_2), 
\label{eq:B5} 
\end{equation} 
where the four parameters $\gamma_2$, $\beta_2$, $A$ and $B$ were fitted 
to the three pieces of data available, namely, the position and width of 
the $\Sigma(1385)$ resonance and the branching ratio for its two main 
decay modes. A family of such parameters is given in Table~\ref{tab:piY}, 
for a range of $A$ values such that the spatial size (here $r_0^{(\pi Y)}$) 
associated with the resulting $\pi Y$ form factors is related physically 
to the spatial size $r_0^{(\pi N)}$ associated with the $P_{33}$ $\pi N$ 
form factor of Table~\ref{tab:piN}. For more details and discussion, 
see Ref.~\cite{GAGA11}.{\footnote{We note that the superscripts $\Lambda$ 
and $\Sigma$ are erroneously interchanged in Eq.~(7) of the published 
journal version where they appear as subscripts. None of the results in 
Ref.~\cite{GAGA11} is affected by this typo.}}

\begin{table}[ht]  
\caption{Fitted parameters of the $\pi\Lambda-\pi\Sigma$ $p$-wave separable 
interaction defined by Eqs.~(\ref{eq:B2}) and (\ref{eq:B5}), for chosen 
values of the parameter $A$. Listed also are values of the r.m.s. 
momentum $\sqrt{<p^2>_{g_2}}$ (in fm$^{-1}$), the r.m.s. radius 
$\sqrt{<r^2>_{\tilde{g_2}}}$ (whenever real) and zero $r_0^{(\pi Y)}$ 
(both in fm) of the Fourier transform ${\tilde {g_2}}(r)$.} 
\begin{ruledtabular} 
\begin{tabular}{ccccccc} 
$A~({\rm fm}^2)$ & $\gamma_2~({\rm fm}^{4})$ & $\beta_2~({\rm fm}^{-1})$ & 
$B$ & $\sqrt{<p^2>_{g_2}}$ & $\sqrt{<r^2>_{\tilde{g_2}}}$ & $r_0^{(\pi Y)}$ \\
\hline 
0.25 & $-$0.0091851  & 2.5810  & 0.93671 & 4.30 & 0.33 & 1.36 \\ 
0.30 & $-$0.0090934  & 2.4765  & 0.95132 & 4.13 & 0.23 & 1.41 \\ 
0.35 & $-$0.0089513  & 2.3919  & 0.96559 & 4.00 &  --  & 1.45 \\ 
0.40 & $-$0.0087763  & 2.3216  & 0.97949 & 3.89 &  --  & 1.48 \\ 
0.45 & $-$0.0085787  & 2.2619  & 0.99298 & 3.80 &  --  & 1.51 \\ 
\end{tabular} 
\end{ruledtabular} 
\label{tab:piY}
\end{table}

\subsection{The $YN$ subsystem} 
\label{sec:YN} 

In the case of isospin $\frac{1}{2}$ which corresponds to the coupled 
$\Lambda N-\Sigma N$ subsystem we have two coupled channels so that 
applying Eq.~(\ref{eq:B1}) to the separable potential 
\begin{equation}
V_3^{YY'}(p_3,p_3^\prime)=\gamma_3^{YY'} g_3^Y(p_3) g_3^{Y'}(p_3^\prime) 
\label{eq:C1} 
\end{equation} 
leads to 
\begin{equation} 
t_3^{YY'}(p_3,p_3^\prime;\omega_0)=
g_3^Y(p_3)\tau_3^{YY'}(\omega_0)g_3^{Y'}(p_3^\prime), 
\label{eq:C2} 
\end{equation} 
where $\tau_3^{YY'}(\omega_0)$ are easily obtained. 
We used Yamaguchi form factors 
\begin{equation} 
g_3^Y(p_3) =\frac{1}{1+(p_3/\alpha_3^Y)^2}, 
\label{eq:C3} 
\end{equation} 
so that there are five free parameters, three strengths and two ranges. 
These five parameters were fitted to the $\Lambda N$ $S=1$ scattering length 
$a_{\frac{1}{2}1}=1.41$ fm and effective range $r_{\frac{1}{2}1}=3.36$ fm, 
the real and imaginary parts of the $\Sigma N$ $S=1$ scattering length 
$a_{\frac{1}{2}1}^\prime=2.74+{\rm i}1.22$ fm, and the phase of the 
$\Lambda N-\Sigma N$ $S=1$ transition scattering length $\psi=23.8^\circ$ 
obtained in the chiral quark model \cite{SALA07}. These parameters are 
given in Table~\ref{tab:YN}. 

\begin{table}[ht]  
\caption{Parameters of the spin-triplet $YN$ separable potentials 
defined by Eqs.~(\ref{eq:C1}) and (\ref{eq:C3}) for isospin values 
$I_{YN}=\frac{1}{2},\frac{3}{2}$.} 
\begin{ruledtabular} 
\begin{tabular}{cccccc} 
$I_{YN}$ & $\gamma_3^{\Lambda\Lambda}$ (fm$^2$) 
& $\gamma_3^{\Lambda\Sigma}$ (fm$^2$) 
& $\gamma_3^{\Sigma\Sigma}$ (fm$^2$) 
& $\alpha_3^\Lambda$ (fm$^{-1}$) &   
$\alpha_3^\Sigma$ (fm$^{-1}$)  \\ 
\hline 
1/2 & $-$0.37704 & $-$0.047865 & $-$0.0059699  &  1.46 & 0.4  \\ 
3/2 &    --      &      --     &   0.36416   &  -- &  1.491   \\ 
\end{tabular} 
\end{ruledtabular}  
\label{tab:YN} 
\end{table} 

The spin-triplet hyperon-nucleon subsystem with isospin $\frac{3}{2}$ 
corresponds to pure $\Sigma N$ scattering and it requires only two free 
parameters, one strength and one range. These two parameters were fitted 
to the $\Sigma N$ $S=1$ scattering length $a_{\frac{3}{2}1}^\prime=-0.44$ 
fm and effective range $r_{\frac{3}{2}1}^\prime=-2.09$ fm obtained in the 
chiral quark model \cite{SALA07}. These parameters are also given in 
Table~\ref{tab:YN}.

\subsection{Compact form of the two-body amplitudes} 
\label{sec:compact} 

The two-body amplitudes discussed above can be written in compact form as 
\begin{eqnarray} 
t_1^Y &=& |g_1^{\pi N}\rangle \tau_1^Y \langle g_1^{\pi N}|, \,\,\,\,\,\, 
Y=\Lambda,\Sigma,  
\label{eq:D1} 
\end{eqnarray} 
\begin{eqnarray} 
t_2 &=& \begin{pmatrix}|g_2^{\pi\Lambda}\rangle\cr |g_2^{\pi\Sigma}\rangle \cr 
\end{pmatrix}\tau_2 \begin{pmatrix} \langle g_2^{\pi\Lambda}| 
 & \langle g_2^{\pi\Sigma}| \cr \end{pmatrix}, 
\label{eq:D2} 
\end{eqnarray} 
\begin{eqnarray} 
t_3 &=& \begin{pmatrix} 
|g_3^{\Lambda N}\rangle\tau_3^{\Lambda N\to\Lambda N}\langle g_3^{\Lambda N}|& 
|g_3^{\Lambda N}\rangle\tau_3^{\Lambda N\to\Sigma N}\langle g_3^{\Sigma N}|\cr 
|g_3^{\Sigma N}\rangle\tau_3^{\Sigma N\to\Lambda N}\langle g_3^{\Lambda N}| & 
|g_3^{\Sigma N}\rangle\tau_3^{\Sigma N\to\Sigma N}\langle g_3^{\Sigma N}| \cr 
\end{pmatrix}. 
\label{eq:D3} 
\end{eqnarray} 
For applications wishing to extend the system of two-body $\pi Y$ coupled 
channels into a system of $\pi Y - \bar K N$ channels, coupled through the 
$\Sigma(1385)$ isobar, Eq.~(\ref{eq:D2}) is to be replaced by 
\begin{eqnarray} 
t_2 &=& \begin{pmatrix}|g_2^{\pi\Lambda}\rangle \cr |g_2^{\pi\Sigma}\rangle 
\cr |g_2^{\bar K N}\rangle \cr \end{pmatrix} \tau_2 \begin{pmatrix} 
\langle g_2^{\pi\Lambda}| & \langle g_2^{\pi\Sigma}| & \langle g_2^{\bar K N}| 
\cr \end{pmatrix}.  
\label{eq:D2'} 
\end{eqnarray}

\section{Three-body equations} 
\label{sec:3body}

Normally, the Faddeev amplitudes are labeled by the spectator particle 
which in general has the same label as the interacting pair. However, 
when there is particle conversion as in the present case one can have 
different interacting pairs for the same spectator or different spectators 
for the same interacting pair. For example, whereas $\pi N$ is the interacting 
pair in the amplitude $T_1$ and the spectator is either $\Lambda$ or $\Sigma$, 
the interacting pair in the amplitude $T_2$ is either $\pi\Lambda$ or 
$\pi\Sigma$ and the spectator is a nucleon. Thus, we will label the 
corresponding Faddeev amplitudes either by the spectator or by the interacting 
pair as helpful as to make the notation clear. In this way, considering all 
possible transitions, one obtains the Faddeev equations 
\begin{equation} 
T_1^Y = t_1^YG_0(\pi YN)T_2^{\pi Y}+t_1^YG_0(\pi YN)T_3^{YN}, 
\label{eq:3body1} 
\end{equation} 
\begin{equation} 
T_2^{\pi Y}= \sum_{Y'} 
t_2^{\pi Y\to\pi Y'}G_0(\pi Y'N)T_1^{Y'}+\sum_{Y'}t_2^{\pi Y\to\pi Y'}
G_0(\pi Y'N)T_3^{Y'N}, 
\label{eq:3body2} 
\end{equation} 
\begin{equation} 
T_3^{YN}=\sum_{Y'}t_3^{YN\to Y'N}G_0(\pi Y'N)T_2^{\pi Y'}
+\sum_{Y'}t_3^{YN\to Y'N}G_0(\pi Y'N)T_1^{Y'}. 
\label{eq:3body3} 
\end{equation} 
For applications wishing to extend the two-body $\pi Y$ coupled channels into 
a system of $\pi Y - \bar K N$ channels coupled through the $\Sigma(1385)$ 
isobar, the Faddeev amplitude (\ref{eq:3body2}) acquires the additional term 
$t_2^{\pi Y\to\bar K N}G_0(\bar K NN)T_2^{\bar K N}$ on the r.h.s., where 
\begin{eqnarray} 
T_2^{\bar K N}&=&t_2^{\bar K N\to\bar K N}G_0(\bar K NN)T_2^{\bar K N} 
 \nonumber \\ 
&+&\sum_{Y}t_2^{\bar K N\to\pi Y}G_0(\pi YN)T_1^{Y}
+\sum_{Y}t_2^{\bar K N\to\pi Y}G_0(\pi YN)T_3^{YN}. 
\label{eq:3body2'} 
\end{eqnarray} 

If we substitute Eq.~(\ref{eq:3body3}) into Eqs.~(\ref{eq:3body1}) and 
(\ref{eq:3body2}), 
using the expressions for the two-body amplitudes (\ref{eq:D1})--(\ref{eq:D3}), 
we get that 
\begin{equation} 
T_1^Y=|g_1^{\pi N}\rangle X_1^Y, \;\;\;\;\;\;\; 
T_2^{\pi Y} = |g_2^{\pi Y}\rangle X_2, 
\label{eq:3body4} 
\end{equation} 
where the new amplitudes $X_1^Y$ and $X_2$ satisfy the equations 
\begin{eqnarray} 
X_1^Y &=& \tau_1^Y\langle g_1^{\pi N}|G_0(\pi YN)|g_2^{\pi Y}\rangle X_2 
 \nonumber \\ 
&+& \sum_{Y^\prime Y^{\prime\prime}}\tau_1^Y\langle g_1^{\pi N}|G_0(\pi Y'N)| 
g_3^{Y'N}\rangle\tau_3^{Y'N\to Y^{\prime\prime}N}\langle 
g_3^{Y^{\prime\prime}N}|G_0(\pi Y^{\prime\prime}N)|g_2^{\pi Y^{\prime\prime}}
\rangle X_2  \nonumber \\ 
&+& \sum_{Y^\prime Y^{\prime\prime}}\tau_1^Y\langle g_1^{\pi N}|G_0(\pi Y'N)| 
g_3^{Y'N}\rangle\tau_3^{Y'N\to Y^{\prime\prime}N}\langle 
g_3^{Y^{\prime\prime}N}|G_0(\pi Y^{\prime\prime}N)|g_1^{\pi N}\rangle 
X_1^{Y^{\prime\prime}}, 
\label{eq:3body5} 
\end{eqnarray} 
\begin{eqnarray} 
X_2 &=& \sum_Y\tau_2\langle g_2^{\pi Y}|G_0(\pi YN)|g_1^{\pi N}\rangle 
X_1^Y  \nonumber \\ 
&+& \sum_{YY'}\tau_2\langle g_2^{\pi Y}|G_0(\pi YN)|g_3^{YN}\rangle
\tau_3^{YN\to Y'N}\langle g_3^{Y'N}|G_0(\pi Y^{\prime}N)|g_2^{\pi Y'}\rangle 
X_2  \nonumber \\ 
&+& \sum_{YY'}\tau_2\langle g_2^{\pi Y}|G_0(\pi YN)|g_3^{YN}\rangle
\tau_3^{YN\to Y'N}\langle g_3^{Y'N}|G_0(\pi Y^{\prime}N)|g_1^{\pi N}\rangle 
X_1^{Y'}. 
\label{eq:3body6} 
\end{eqnarray} 
As shown in Ref.~\cite{GAGA11}, the one-dimensional integral equations 
corresponding to the Faddeev equations for the $\pi\Lambda N-\pi\Sigma N$ 
system can be read off from the AGS form Eqs.~(\ref{eq:3body5}) and 
(\ref{eq:3body6}). 

For applications wishing to extend the description of the $\Sigma(1385)$ 
isobar in terms of $\pi Y$ coupled channels into $\pi Y - \bar K N$ coupled 
channels, the definition of $X_2$ in Eq.~(\ref{eq:3body4}) is generalized to 
\begin{eqnarray} 
\begin{pmatrix}T_2^{\pi Y}\cr T_2^{\bar K N}\cr \end{pmatrix} &=& 
\begin{pmatrix}|g_2^{\pi Y}\rangle\cr |g_2^{\bar K N}\rangle\cr \end{pmatrix}
X_2, 
\label{eq:3body4'} 
\end{eqnarray} 
with Eq.~(\ref{eq:3body6}) modified by adding on its r.h.s. the term 
$\tau_2\langle g_2^{\bar K N}|G_0(\bar K NN)|g_2^{\bar K N}\rangle X_2$.

\section{Results} 
\label{sec:res} 

We started by searching for ($I=3/2,J^P=2^+$) $\pi\Lambda N-\pi\Sigma N$ 
bound-state poles, i.e. considering real values of $W_0<m_\pi+m_\Lambda+m_N$ 
for which there are no three-body singularities. The one-dimensional 
integral equations which follow from the coupled-amplitude AGS 
equations (\ref{eq:3body5}) and (\ref{eq:3body6}) were solved. 
Unlike the nonrelativistic cases studied in \cite{GAGA08} and \cite{GAGA10} 
we found no pole which would correspond to a bound state. 
In order to artificially generate such a pole we multiplied the strengths 
$\gamma_1$ and $\gamma_2$ by factors $f_1>1$ and $f_2>1$ which exactly 
produce a bound state pole at the $\pi\Lambda N$ threshold 
$W_0=m_\pi+m_\Lambda+m_N$. We then rotated the integration contour 
into the complex plane as described in \cite{GAGA11}, i.e., 
$q_i\to q_i\exp(-{\rm i}\phi)$ which allowed us to reduce slowly the factors 
$f_i$ and follow the bound state pole into the complex plane to its final 
position once $f_1=f_2=1$. Finally, we checked that the position 
of the pole is independent of the value of $\phi$. 

\begin{table}[ht] 
\caption{Energy position of the $\pi\Lambda N$ resonance pole, relative 
to the $\pi\Sigma N$ threshold, calculated for the $g_2$ form factors 
of Table~\ref{tab:piY}, listed according to their $A$ parameter and the 
zero of ${\tilde{g_2}}$.} 
\begin{ruledtabular} 
\begin{tabular}{ccccccc} 
& & $A~({\rm fm}^2)$ & $r_0^{(\pi Y)}~({\rm fm})$ & $E~({\rm MeV})$ & & \\
\hline  
& & 0.25 & 1.36 & $-$19.8$-$i2.6 & & \\ 
& & 0.30 & 1.41 & $-$17.6$-$i2.9 & & \\ 
& & 0.35 & 1.45 & $-$15.6$-$i3.2 & & \\ 
& & 0.40 & 1.48 & $-$13.7$-$i3.5 & & \\ 
& & 0.45 & 1.51 & $-$11.9$-$i3.8 & & \\  
\end{tabular} 
\end{ruledtabular} 
\label{tab:res} 
\end{table} 

In Table~\ref{tab:res} we list the energy eigenvalues, measured with respect 
to the $\pi\Sigma N$ threshold, as calculated using the $P_{33}$ $\pi N$ 
form factor from Table~\ref{tab:piN} and the family of $\pi Y$ form factors 
recorded in Table~\ref{tab:piY}. The sensitivity of the calculated pole 
energy to the parametrization of the $\pi Y$ form factor amounts to less 
than 10 MeV. In all cases the eigenvalue lies above the $\pi\Lambda N$ 
threshold, but below the $\pi\Sigma N$ threshold. If we neglect the $YN$ 
interaction, the real part of the pole energy rises approximately 10 MeV 
while the imaginary part remains almost the same. Finally, in order to check 
the effect of other non-resonating partial waves, we repeated the calculation 
of the first row in Table~\ref{tab:res} adding the $\pi N$ $P_{13}$ partial 
wave from the second row of Table~\ref{tab:piN}. The energy changed then from 
$E=-19.755-{\rm i}2.611$ MeV to $E=-19.734-{\rm i}2.613$ MeV, demonstrating 
that this effect is quite negligible.

\section{Discussion} 
\label{sec:disc} 

In this section we discuss two aspects of the present relativistic 
three-body calculation, (i) relativistic vs nonrelativistic and (ii) 
the inclusion of a $\bar K NN$ channel.

\subsection{Relativistic vs Nonrelativistic} 
\label{sec:rel} 

As observed in the previous section the effects of a relativistic 
treatment are quite important for the $\pi\Lambda N-\pi\Sigma N$ system, 
removing the $\pi\Lambda N$ bound-state solution obtained in the 
nonrelativistic (NR) model \cite{GAGA08,GAGA10}. 

In order to understand the origin of the discrepancy between the relativistic 
and NR results we have repeated the calculation of the $\pi\Lambda N$ 
problem \cite{GAGA08} for the simple case where there is no coupling to 
the $\pi\Sigma N$ channel and one neglects the $YN$ interaction. In this case, 
the Faddeev equations of the $\pi\Lambda N$ bound-state problem are 
\begin{eqnarray} 
X_{\pi N} &=& \tau_{\pi N}\langle g_{\pi N}|G_0(\pi\Lambda N)
|g_{\pi\Lambda}\rangle X_{\pi\Lambda}, 
\label{eq:discA1} \\ 
X_{\pi\Lambda} &=& \tau_{\pi\Lambda}\langle g_{\pi\Lambda}|G_0(\pi\Lambda N) 
|g_{\pi N}\rangle X_{\pi N}, 
\label{eq:discA2} 
\end{eqnarray} 
where $\tau_{\pi i}$ with $i$=$N,\Lambda$ are the isobar propagators of the 
$\pi i$ subsystems and $\langle g_{\pi i}|G_0(\pi\Lambda N)|g_{\pi j}\rangle$ 
are the one-pion-exchange diagrams. The $\pi N$ and $\pi\Lambda$ separable 
potentials used in \cite{GAGA08} are of the form 
\begin{equation} 
V_{\pi i}(p,p')=\gamma_{\pi i}g_{\pi i}(p)g_{\pi i}(p'), 
\label{eq:discA3} 
\end{equation} 
with 
\begin{equation} 
g_{\pi i}(p)=p(1+p^2)\exp(-p^2/\alpha_{\pi i}^2), 
\label{eq:discA4} 
\end{equation} 
where the parameters $\gamma_{\pi i}$ and $\alpha_{\pi i}$ were fitted 
to the position and width of the resonances as given by the Particle Data 
Group \cite{PDG}. We list these parameters in Table~\ref{tab:discA} as well 
as the corresponding ones obtained using the relativistic formulation in 
Ref.~\cite{GAGA11}. Using the parameters listed in the table, the NR model 
predicts a bound state at about $-110$ MeV while in the case of the 
relativistic model there is no bound state. If in the relativistic model we 
replace the one-pion-exchange diagrams by their NR versions we obtain almost 
the same results for the Fredholm determinant and consequently no bound state. 
On the other hand, if we replace the isobar propagators by their NR versions, 
the Fredholm determinant changes radically giving rise to even deeper bound 
state. Thus, the problem with the NR model lies in the definition of the 
isobar propagators. 

\begin{table}[ht] 
\caption{Parameters of the $\pi N$ and $\pi\Lambda$ separable potentials 
Eqs.~(\ref{eq:discA3}) and (\ref{eq:discA4}) for the nonrelativistic (NR) 
and relativistic (R) models as well as the corresponding isobar 
propagators evaluated at $W_0=m_\pi+m_\Lambda+m_N$ and $q_i=0$.} 
\begin{ruledtabular} 
\begin{tabular}{ccccccc} 
Model & $\gamma_{\pi N}$ (fm$^2$) 
& $\alpha_{\pi N}$ (fm$^{-1}$) & $\tau_{\pi N}(W_0;q_i)$ (fm$^2$) 
& $\gamma_{\pi\Lambda}$ (fm$^2$) 
& $\alpha_{\pi\Lambda}$ (fm$^{-1}$) &  
$\tau_{\pi\Lambda}(W_0;q_i)$ (fm$^2$)  \\ 
\hline 
NR & $-$0.02116  & 2.02135  & $-$0.091220  & $-$0.00564  & 2.523999  
& $-$0.042807  \\ 
R & $-$0.01463  & 1.85836  & $-$0.035758  & $-$0.00471  & 2.236443 
& $-$0.016387  \\ 
\end{tabular} 
\end{ruledtabular} 
\label{tab:discA} 
\end{table} 

The isobar propagators of the relativistic model are given by 
Eq.~(\ref{eq:A7}) of this paper, while the NR ones are given by 
\begin{equation} 
[\tau_{\pi i}(W_0,q_i)]^{-1} = \frac{1}{\gamma_{\pi i}}-\int_0^\infty 
p_i^2 dp_i \frac{g_{\pi i}^2(p_i)}{W_0-m_\pi-m_\Lambda-m_N-p_i^2/2\eta_i
-q_i^2/2\nu_i+{\rm i}\epsilon}, 
\label{eq:discA5} 
\end{equation} 
where $\eta_i$ and $\nu_i$ are the usual reduced masses. We give in the 
table the value of the isobar propagators of the NR and relativistic models 
for $W_0=m_\pi +m_\Lambda +m_N$ and $q_i=0$. As one sees, the NR isobar 
propagators are about three times larger than the relativistic ones. 
In addition, from Eqs.~(\ref{eq:A7}) and (\ref{eq:discA5}) one sees that 
$\tau_{\pi i}(W_0,q_i) \to \gamma_{\pi i}$ when $q_i\to\infty$, so that from 
the values of Table~\ref{tab:discA} one sees that also in this limit the 
NR isobar propagators are larger than the relativistic ones and hence 
artificially boost the attraction, thereby giving rise to the appearance 
of bound states in the case of a NR theory. 

The large differences between the nonrelativistic and relativistic isobar 
propagators can be understood by observing that the $\pi N$ $\Delta(1232)$ 
resonance is 154 MeV above the $\pi N$ threshold and the $\pi\Lambda$ 
$\Sigma(1385)$ resonance is 131 MeV above the $\pi\Lambda$ threshold, 
i.e., the excitation energies are approximately equal to the mass of the 
pion and therefore the use of nonrelativistic kinematics is not appropriate. 

In Ref.~\cite{GAGA08} we also presented results based in the relativistic 
on-mass-shell spectator formalism \cite{GROSS82,GARC87,STAD97} which produced 
similar bound states as the nonrelativistic formalism. We checked that the 
problem here lies again in the isobar propagators even though the kinematics 
is relativistic. The problem, as we pointed out in \cite {GAGA08}, is that 
solutions that fit the experimental data exist only if one puts the light 
particle (in this case the pion) on the mass shell while physically one 
expects that rather the heavy particle ($N$ or $\Lambda$) should be the one 
staying on the mass shell.

\subsection{Including ${\bar K}NN$} 
\label{sec:K} 

Here we study the effects of expanding the three-body model 
space from $\pi\Lambda N-\pi\Sigma N$ coupled channels to 
$\pi\Lambda N-\pi\Sigma N-{\bar K}NN$ coupled channels. The primary reason 
to exclude the ${\bar K}NN$ channel from the very beginning was that the 
three-body quantum numbers $I=\frac{3}{2},J^P=2^+$ are compatible only 
with a Pauli forbidden $I_{NN}=1, J^P=1^+$ leading $NN$ configuration. 
A secondary reason was that although SU(3) predicts a natural-size coupling 
between the $\bar KN$ and $\pi Y$ two-body channels through the $\Sigma(1385)$ 
$p$-wave resonance, there is ample empirical evidence that this coupling 
is quite weak \cite{Kim67,Martin81,CFGGM11}. To extend the relativistic 
$\pi\Lambda N-\pi\Sigma N$ coupled channels calculation, we generalized 
the $\pi Y$ form factors (\ref{eq:B5}) to include also a coupled $\bar K N$ 
form factor as follows: 
\begin{equation} 
g_2^{\Lambda}(p_2)=p_2(1+Ap_2^2)\exp(-p_2^2/\beta_2^2),\,\,\, 
g_2^{\Sigma}(p_2)=Bg_2^{\Lambda}(p_2),\,\,\, 
g_2^{\bar K N}(p_2)=Cg_2^{\Lambda}(p_2),
\label{eq:discB1} 
\end{equation} 
with an overall strength parameter $\gamma_2$. 
The fitted parameters, starting with the parameters in the first row of 
Table~\ref{tab:piY} for $C=0$ and varying $C$ between 0 to 1, are listed 
in Table~\ref{tab:KbarN} together with the pole energy with respect to the 
$\pi\Sigma N$ threshold as obtained by solving the one-dimensional integral 
equations corresponding to the Faddeev equations in the AGS form given by 
Eqs.~(\ref{eq:3body5}) and (\ref{eq:3body6}), with the modification indicated 
at the end of section \ref{sec:3body}. 

\begin{table}[ht] 
\caption{Fitted parameters of the $\pi Y - \bar K N$ form factors 
(\ref{eq:discB1}), for $A=0.25$ and a sequence of values $C=0\cdots 1$, 
together with pole energies with respect to the $\pi\Sigma N$ threshold 
obtained by solving the three-body equations.} 
\begin{ruledtabular} 
\begin{tabular}{ccccc} 
$C$ & $\beta_2$~(fm$^{-1}$) & $\gamma_2$~(fm$^4$) & $B$ & $E$~(MeV)  \\  
\hline 
0.0 & 2.5810 & $-$0.0091851 & 0.9367 & $-$19.8$-$i2.6  \\
0.1 & 2.5774 & $-$0.0092150 & 0.9364 & $-$18.7$-$i2.8  \\ 
0.2 & 2.5668 & $-$0.0093005 & 0.9356 & $-$15.6$-$i3.2  \\
0.3 & 2.5497 & $-$0.0094420 & 0.9342 & $-$10.9$-$i4.0  \\
0.4 & 2.5264 & $-$0.0096400 & 0.9323 & $-$5.0$-$i5.2  \\
0.5 & 2.4978 & $-$0.0098901 & 0.9299 & $+$1.8$-$i6.9  \\
0.6 & 2.4646 & $-$0.0101955 & 0.9269 & $+$8.8$-$i9.1  \\
0.7 & 2.4276 & $-$0.0105512 & 0.9236 & $+$15.7$-$i11.6  \\
0.8 & 2.3876 & $-$0.0109590 & 0.9197 & $+$22.2$-$i14.6  \\
0.9 & 2.3452 & $-$0.0114181 & 0.9155 & $+$27.9$-$i17.8  \\
1.0 & 2.3011 & $-$0.0119291 & 0.9108 & $+$33.0$-$i21.2  \\
\end{tabular} 
\end{ruledtabular} 
\label{tab:KbarN} 
\end{table} 

It is seen that the $Y=1,I=\frac{3}{2},J^P=2^+$ resonance energy goes up 
monotonically upon boosting the $\bar K N-\pi Y$ coupling via increasing 
the parameter $C$. For weak coupling the resonance energy is still below 
the $\pi\Sigma N$ threshold, but for strong coupling ($C\geq 0.5$) it is 
above this threshold. Altogether, the variation of the real part of the 
energy amounts to about 50 MeV upward shift for $C$ between 0 to 1. 
This is accompanied by a substantial increase of the width from about 5 to 
40 MeV. We estimate $C \lesssim 0.2$ from studies of $\Sigma(1385)$ impact 
on low-energy and subthreshold $\bar K$-nucleon \cite{Kim67,Martin81} and 
$\bar K$-nucleus \cite{CFGGM11} phenomenology. Hence, it is fair to conclude 
that the effect of including explicitly a weakly coupled $\bar K NN$ channel 
in the present $\pi\Lambda N-\pi\Sigma N$ coupled channels calculation is 
rather insignificant.

\section{Summary and conclusions} 
\label{sec:sum} 

In this work we have formulated and solved a set of relativistic three-body 
Faddeev equations for $\pi\Lambda N - \pi\Sigma N$ coupled channels in search 
for a bound state or a resonance with quantum numbers $I=3/2, J^P=2^+$. 
The leading two-body attractive interactions were $p$-wave 
interactions in the $\pi N$ and $\pi\Lambda-\pi\Sigma$ channels dominated 
by the $\Delta(1232)$ and $\Sigma(1385)$ resonances, respectively, and to 
a lesser extent the $^{3}S_1$ $YN$ $s$-wave interactions. These interactions 
were fitted by energy-independent separable forms constrained by available 
data. In particular, the $\Delta(1232)$ and $\Sigma(1385)$ members of the 
SU(3) baryon decuplet were generated dynamically as $p$-wave meson-baryon 
resonances without recourse to their intrinsic quark structure. A robust 
$\pi\Lambda N$ resonance some 10--20 MeV below the $\pi\Sigma N$ threshold 
was found upon solving the relativistic three-body coupled channels equations. 
This prediction outdates our earlier prediction of a $\pi\Lambda N$ bound 
state \cite{GAGA08,GAGA10} which was based on a nonrelativistic formulation 
shown here to be inappropriate. Also discussed in the present work was the 
effect of coupling a $\bar K NN$ channel to the $\pi\Lambda N-\pi\Sigma N$ 
driving channels, which turned out to be a secondary effect. 

We conjecture that the ($I=3/2,J^P=2^+$) $\pi\Lambda N$ resonance calculated 
in the present work provides the lowest-mass strangeness ${\cal S}=-1$ 
$s$-wave dibaryon which we denote $\cal Y$. It may be viewed as a $^5S_2$ 
$\Sigma(1385)N-\Delta(1232)Y$ quasibound state with mass $M({\cal Y})$ over 
50 MeV below the lowest threshold ($\Sigma(1385)N$) and over 150 MeV below 
the ($I=1/2,J^P=2^+$) $\Sigma(1385)N-\Delta(1232)\Sigma$ 
dibaryon configuration which provides the lowest ${\cal S}=-1$ dibaryon 
predicted in quark-gluon dynamics \cite{Oka88}. In the present underlying 
meson-baryon dynamics, with pion assisted dibaryons, the ($I=1/2,J^P=2^+$) 
$\Sigma(1385)N-\Delta(1232)\Sigma$ configuration is realized as 
a three-body $\pi\Lambda N$ resonance at $E=90-{\rm i}52$ MeV with respect 
to the $\pi\Sigma N$ threshold, for the same two-body interactions that 
produce the ($I=3/2,J^P=2^+$) $\pi\Lambda N$ resonance at $E=-20-{\rm i}2.6$ 
MeV (first row, Table~\ref{tab:res}). This difference of about 100 MeV 
arises because the $p$-wave $\pi B$ interactions in the $I=1/2$ three-body 
configuration are no longer completely exhausted by the resonating 
$\Delta(1232)$ and $\Sigma(1385)$ isobars. 

The ($I=3/2,J^P=2^+$) $\pi\Lambda N$ resonance found in this work is rather 
narrow. Its `fall-apart' width is seen from Table~\ref{tab:KbarN} to increase 
from a few MeV to over 40 MeV as the resonance energy goes up by about 50 MeV. 
Extrapolating Im~$E$ as a function of Re~$E$, a width of 113 MeV is obtained 
for Re~$E=76$ MeV, this latter value providing the excitation energy of 
$\Delta(1232)$ with respect to the two-body $\pi N$ system, assuming the 
$\Lambda$ hyperon is at rest. This width is (perhaps fortuitously) close to 
the free-space $\Delta$ width of 110 MeV deduced from the input $P_{33}$ phase 
shifts.  

The small `fall-apart' width does not include the effect of true pion 
absorption into a $d$-wave $\Sigma N$ lower channel which was disregarded 
in the present work. Further calculations are necessary to clarify the effect 
of incorporating this pionless channel in our three-body formulation, but its 
inclusion is unlikely to disrupt the existence of the $\pi\Lambda N$ resonance 
explored here. We note that a $d$-wave $\Sigma N$ configuration is connected 
by a strong one-pion exchange (OPE) tensor potential to the $^5S_2$ 
$\Sigma(1385)N$ and $\Delta(1232)Y$ components of the dibaryon $\cal Y$. 
Such OPE tensor transition potential could give rise to a pionless decay 
width of $\cal Y$ in the range of few tens of MeV, employing estimates similar 
to those made for the width of quasibound $\Sigma$ hyperon nuclear states 
arising from the $\Sigma N({^{3}S}_1)\to\Lambda N({^{3}D}_1)$ OPE tensor 
transition potential \cite{GD80}. 

The structure of $\cal Y$ is reminiscent of the ${\cal S}=0$ ($I=0,{^{7}S}_3$) 
$s$-wave $\Delta\Delta$ dibaryon candidate recently observed in double-pion 
production reactions in $NN$ collisions \cite{clement12}. The $\cal Y$ 
dibaryon could also be searched in $pp$ collisions, say in 
\begin{eqnarray} 
   p~ + ~p & ~\rightarrow ~ & {\cal Y}^{++} ~+~K^0 \nonumber  \\  
           &                & ~\hookrightarrow ~ \Sigma^+ ~+~ p 
\label{eq:pptoY++} 
\end{eqnarray} 
at energies above the $\Sigma(1385)$ production threshold. Here, owing to the 
doubly-positive charge $Q=+2$, the decay ${\cal Y}^{++} \to \Sigma^+ p$ offers 
a unique decay channel. The production and decay (\ref{eq:pptoY++}) 
are analogous to those conjectured for the ($Y=1,I=1/2,J^P=0^-$) $\bar KNN$ 
quasibound state $\cal K$ in the recent DISTO re-analysis at 
$T_p=2.85$ GeV \cite{DISTO10}: 
\begin{eqnarray} 
   p~ + ~p & ~\rightarrow ~ & {\cal K}^+ ~+~K^+ \nonumber  \\  
           &                & ~\hookrightarrow ~ \Lambda ~+~ p. 
\label{eq:pptoX+} 
\end{eqnarray} 
Of course, ${\cal Y}$ may also be studied in $pp$ collisions with outgoing 
$K^+$ meson, but the decay ${\cal Y}^+ \to (\Sigma^+ n,\;\Sigma^0 p)$ may 
not be easily distinguished from the decay ${\cal K}^+\to(\Sigma^+ n,\;
\Sigma^0 p)$. The production of $\Sigma(1385)$ charge states in $pp$ 
collisions with outgoing $K^+$ meson has been studied recently in great 
detail by the HADES Collaboration at GSI \cite{HADES12}. 
 
Other possible production reactions are 
\begin{eqnarray} 
K^- ~+~ d & ~\rightarrow ~ & {\cal Y}^{-} ~+~\pi^+ \nonumber  \\ 
 &  &  ~\hookrightarrow ~ \Sigma^- +n, 
\label{eq:K-dtoY-} 
\end{eqnarray} 
\begin{eqnarray} 
\pi^- ~+~ d & ~\rightarrow ~ & {\cal Y}^{-} ~+~K^+  \nonumber  \\ 
 &  &  ~\hookrightarrow ~ \Sigma^- +n, 
\label{eq:pi-dtoY-} 
\end{eqnarray} 
\begin{eqnarray} 
\pi^+ ~+~ d & ~\rightarrow ~ & {\cal Y}^{++} ~+~K^0  \nonumber  \\ 
 &  &  ~\hookrightarrow ~ \Sigma^+ +p, 
\label{eq:pi+dtoY++} 
\end{eqnarray} 
or 
\begin{eqnarray} 
\pi^+ ~+~ d & ~\rightarrow ~ & {\cal Y}^{+} ~+~K^+  \nonumber  \\ 
  &  & ~\hookrightarrow ~ \Sigma^+ +n,\; \Sigma^0 +p,  
\label{eq:pi+dtoY+} 
\end{eqnarray} 
similar to the E27 experiment scheduled at J-PARC \cite{NagaeE27}: 
\begin{eqnarray} 
\pi^+ ~+~ d & ~\rightarrow ~ & {\cal K}^+ ~+~K^+ \nonumber  \\  
          &                & ~\hookrightarrow ~ \Lambda ~+~ p. 
\label{eq:pi+dtoX} 
\end{eqnarray} 
This structural similarity between production and decay schemes of $\cal K$ 
and of $\cal Y$ helps to realize that the proposed ($I=3/2,J^P=2^+$) $\cal Y$ 
dibaryon is related to a dominant $\Sigma(1385)N$ configuration much the same 
as the ($I=1/2,J^P=0^-$) $\cal K$ dibaryon is related to a dominant 
$\Lambda(1405)N$ configuration. For both dibaryons, pionic three-body decay 
modes, ${\cal K}\to \pi\Sigma N$ and ${\cal Y}\to \pi\Lambda N$ may also 
provide useful experimental signature, provided they are energetically 
allowed.

\begin{acknowledgments} 

The research of HG is supported in part by COFAA-IPN (M\'exico) and the 
research of AG by the HadronPhysics3 networks SPHERE and LEANNIS of the 
European FP7 initiative. 

\end{acknowledgments}


\begin{thebibliography}{99} 

%\bibitem{FINUDA} M.~Agnello, {\it et al.}, Phys. Rev. Lett. {\bf 94}, 212303 
%(2005). 

%\bibitem{AKAISH} Y.~Akaishi, and T.~Yamazaki, Phys. Rev. C {\bf 65}, 044005 
%(2002). 

%\bibitem{SHEV1} N.V.~Shevchenko, A.~Gal, and J.~Mare\^s, Phys. Rev. Lett. 
%{\bf 98}, 082301 (2005). 

%\bibitem{SHEV2} N.V.~Shevchenko, A.~Gal, J.~Mare\^s, and J.~R\'evai, Phys. 
%Rev. C {\bf 76}, 044004 (2007). 

%\bibitem{WEISE1} T.~Hyodo, and W.~Weise, Phys. Rev. C {\bf 77}, 035204 
%(2008). 

%\bibitem{WEISE2} A.~Dot\'e, T.~Hyodo, and W.~Weise, Phys. Rev. C {\bf 79}, 
%014003 (2009). 

\bibitem{GAGA08} A.~Gal and H.~Garcilazo, Phys. Rev. D {\bf 78}, 014013 
(2008). 

\bibitem{GAGA10} H.~Garcilazo and A.~Gal, Phys. Rev. C {\bf 81}, 055205 
(2010). 

\bibitem{GAGA11} A.~Gal and H.~Garcilazo, Nucl. Phys. A {\bf 864}, 153 
(2011). 

\bibitem{Ikeda07} Y.~Ikeda and T.~Sato, Phys. Rev. C {\bf 76}, 035203 (2007). 

\bibitem{GAL11} A.~Gal, in {\it From Nuclei to Stars, Festschrift in Honor 
of Gerald E. Brown}, Ed. Sabine Lee (World Scientific, 2011) pp. 157-170 
[arXiv:1011.6332]. 

\bibitem{Oka12} M.~Oka, Nucl. Phys. A {\bf 881}, 6 (2012). 

\bibitem{HM12} J.~Haidenbauer and U.-G.~Mei{\ss}ner, Nucl. Phys. A {\bf 881}, 
44 (2012). 

\bibitem{Ikeda10} Y.~Ikeda, H.~Kamano, and T.~Sato, Prog. Theor. Phys. 
{\bf 124}, 533 (2010). 

\bibitem{SALA07} H.~Garcilazo, T.~Fern\'andez-Caram\'es, and A.~Valcarce, 
Phys. Rev. C {\bf 75}, 034002 (2007). 
 
\bibitem{PDG} J.~Beringer {\it et al.}, Phys. Rev. D {\bf 86}, 010001 (2012). 

\bibitem{GROSS82} F.~Gross, Phys. Rev. C {\bf 26}, 2226 (1982). 

\bibitem{GARC87} H.~Garcilazo, Phys. Rev. C {\bf 35}, 1804 (1987). 

\bibitem{STAD97} A.~Stadler, F.~Gross, and M.~Frank, Phys. Rev. C {\bf 56}, 
2396 (1997). 

\bibitem{Kim67} J.K.~Kim, Phys. Rev. Lett. {\bf 19}, 1074 (1967).

\bibitem{Martin81} A.D.~Martin, Nucl. Phys. B {\bf 179}, 33 (1981).

\bibitem{CFGGM11} A.~Cieply, E.~Friedman, A.~Gal, D.~Gazda, and J.~Mares, 
Phys. Rev. C {\bf 84}, 045206 (2011). 

\bibitem{Oka88} M.~Oka, Phys. Rev. D {\bf 38}, 298 (1988). 

\bibitem{GD80} A.~Gal and C.B.~Dover, Phys. Rev. Lett. {\bf 44}, 379 (1980). 

\bibitem{clement12} H.~Clement, Prog. Part. Nucl. Phys. {\bf 67}, 486 (2012), 
and references cited therein. 
%P.~Adlarson et al., Phys. Rev. Lett. {\bf 106}, 242302 (2011). 

\bibitem{DISTO10} T.~Yamazaki, {\it et al.}, Phys. Rev. Lett. {\bf 104}, 
132502 (2010).

\bibitem{HADES12} G.~Agakishiev, {\it et al.}, Phys. Rev. C {\bf 85}, 035203 
(2012), Nucl. Phys. A {\bf 881}, 178 (2012). 

\bibitem{NagaeE27} http://j-parc.jp/researcher/Hadron/en/Proposal$\_$e.html, 
proposal P27: Search for a nuclear $\bar K$ bound state $K^-pp$ in the 
$d(\pi^+,K^+)$ reaction , T.~Nagae spokesperson.  

\end{thebibliography}
\end{document}